\documentclass{emulateapj}
\usepackage{graphicx,amsmath,color,natbib}

\newcommand{\ngq}{N_g(Q)} \newcommand{\ngg}{N_g(G)}
\newcommand{\ngr}{N_g(R)} \newcommand{\ngl}{N_g(L^*)}
 
 \def\arcsec{$^{\prime \prime}$}

\begin{document}

\title{The Small-Scale Environment of Quasars}

\author{Will~Serber\altaffilmark{1}, Neta~Bahcall\altaffilmark{1},
Brice~M\'{e}nard\altaffilmark{2},
Gordon~Richards\altaffilmark{1}\altaffilmark{3}}
\altaffiltext{1}{Princeton University Observatory, Princeton, NJ
08544, USA} \altaffiltext{2}{Institute for Advanced Study, Einstein
Drive, Princeton, NJ 08540, USA} \altaffiltext{3}{Department of
Physics and Astronomy, The Johns Hopkins University, 3400 North
Charles Street, Baltimore, MD 21218-2686}

\slugcomment{Dec15 '05}

\begin{abstract}

Where do quasars reside? Are quasars located in environments similar
to those of typical $L^*$ galaxies, and, if not, how do they differ?
An answer to this question will help shed light on the triggering
process of quasar activity. We use the Sloan Digital Sky Survey to
study the environment of quasars and compare it directly with the
environment of galaxies. We find that quasars ($M_i \le -22$, $z \le
0.4$) are located in higher local overdensity regions than are typical
$L^*$ galaxies. The enhanced environment around quasars is a local
phenomenon; the overdensity relative to that around $L^*$ galaxies is
strongest within $100\,$kpc of the quasars. In this region, the
overdensity is a factor of 1.4 larger than around $L^*$ galaxies. The
overdensity declines monotonically with scale to nearly unity at
$\sim1\,h_{70}^{-1}\,$Mpc, where quasars inhabit environments
comparable to those of $L^*$ galaxies. The small-scale density
enhancement depends on quasar luminosity, but only at the brightest
end: the most luminous quasars reside in higher local overdensity
regions than do fainter quasars. The mean overdensity around the
brightest quasars ($M_i \le -23.3$) is nearly three times larger than
around $L^*$ galaxies while the density around dimmer quasars ($M_i =
-22.0$ to $-23.3$) is $\sim1.4$ times that of $L^*$ galaxies. By
$\sim0.5\,$Mpc, the dependence on quasar luminosity is no longer
significant. The overdensity on all scales is independent of redshift
to $z \le 0.4$. The results suggest a picture in which quasars
typically reside in $L^*$ galaxies, but have a local excess of
neighbors within $\sim0.1\,-\,0.5\,$Mpc; this local density excess
likely contributes to the triggering of quasar activity through
mergers and other interactions.

\end{abstract}


\section{Introduction} \label{sec.intro}

For more than two decades, significant effort has been spent
attempting to understand the triggering mechanism of quasar activity,
as well as the relation between quasars and their host galaxies.
Since \citet{bell_1969}, it has become widely accepted that quasars
are fueled by accretion of gas onto super-massive black holes.
Observations have shown that a number of nearby galaxies have a
central black hole whose mass correlates with the luminosity of the
spheroid of the host galaxy. This connection suggests that the
formation of the black hole is linked to the formation of the galaxy
which, in turn, is known to strongly depend on its environment. To
develop a better understanding of the quasar phenomenon, it is
therefore important to investigate and quantify the relation between
quasars and their environments. Despite the importance of this issue,
our knowledge of the quasar environment is still limited.

Quasar environments have been studied on different scales ranging from
those of the host galaxy to those of large scales. Such studies have
provided important but controversial results regarding the environment
of quasars. It has been known for more than three decades that quasars
are associated with enhancements in the spatial distribution of
galaxies \citep{bahcall_1969}. Studies have shown that, in the nearby
universe, quasars reside in environments ranging from small to
moderate groups of galaxies rather than in rich clusters
(\citealt{bahcall_1991b,fisher_1996,mclure_2001}).

Early observations of quasar environments \citep{stockton_1978,
yee_1984, yee_1987, boyle_1988, smith_1990, ellingson_1991}, revealed
a positive association of bright quasars with neighboring galaxies at
a level somewhat higher than that of normal galaxies and comparable to
the environment of small- to intermediate-richness groups of galaxies
\citep[e.g.,][]{bahcall_1991}. Observations of quasar environment from
the Hubble Space Telescope snapshot survey \citep{bahcall_1997}
further support this density enhancement around bright quasars. All
these observations focused on small scales, typically within $\sim
0.5\,$Mpc of the quasars, and used relatively small samples of objects.

Early observations of the clustering properties of quasars themselves,
as measured by the quasar auto-correlation function, suggest that
quasars are significantly more strongly clustered than galaxies on
scales up to $10\,$Mpc and greater \citep[e.g.,][]{shaver_1988,
shanks_1988, chu_1988, chu_1989, crampton_1989}, but less clustered
than rich clusters of galaxies \citep[e.g.,][]{bahcall_1991}. This
finding suggests that quasars are located in high overdensity regions,
more so than $L^*$ galaxies, since higher overdensity regions are
clustered more strongly than lower overdensity regions
\citep[e.g.,][]{bahcall_1983, kaiser_1984, bardeen_1986}. An overdense
environment would indeed be expected if the quasar activity was
triggered by galaxy interactions.

On the other hand, new generation surveys, such as the Two Degree
Field (2dF) and the Sloan Digital Sky Survey (SDSS), have given rise
to different results. Using significantly larger complete samples of
quasars and galaxies, these surveys have shown that on large scales,
i.e. from 1 to $10\,$Mpc, the quasar-galaxy cross-correlation and the
quasar auto-correlation are comparable to the correlation function of
$L^*$ galaxies. This suggests, in conflict with previous results, that
quasars and Active Galactic Nuclei (AGNs) inhabit environments similar
to those of $L^*$ galaxies \citep[e.g.,][]{smith_1995, croom_1999,
croom_2003, miller_2003, kauffmann_2004, wake_2004}.  The results
also suggest that the quasar correlation function does not depend
significantly on either quasar luminosity or redshift within the
ranges studied.

Recent work on sub-Mpc scales using the SDSS to find quasar pairs
suggests that the quasar-quasar auto-correlation function may be
enhanced relative to the galaxy-galaxy distribution
\citep{hennawi_2005}, consistent with the earlier results on small
scales, as discussed above.

In this paper we use the SDSS survey to determine the galaxy
environment around quasars as a function of scale. The SDSS is
uniquely suited for this investigation: it is the largest complete
survey available of both galaxies and quasars, carried out in a
well-calibrated, self-consistent manner. The data used in this study
covers 4000 deg$^2$, with $\sim2\times10^3$ quasars of redshift
$z\le0.4$ and ten million photometric galaxies to a magnitude limit of
$i = 21$. We use these data to determine the mean galactic environment
around quasars as a function of quasar luminosity and redshift. For
comparison, the same analysis is then repeated to find the local
environment around $10^5$ spectroscopic galaxies in the SDSS area, as
well as around random positions in the survey. All the analyses are
carried out using the same ten million photometric galaxies. This
technique allows a direct comparison between the environment around
quasars with that around random points as well as with the environment
around $L^*$ galaxies, thereby minimizing potential selection effects and
systematics.

The outline of the paper is as follows: we discuss the data in
Section~\ref{sec.data}, the analysis in Section~\ref{sec.analysis},
and the results in Section~\ref{sec.results}. The conclusions are
summarized in Section~\ref{sec.conclusions}. Throughout this paper, we
use a cosmological model with $H_0\,=\,70\,{\rm km^{-1}\,Mpc^{-1}}$,
$\Omega_M\,=\,0.3$, and $\Omega_{\Lambda}\,=\,0.7$ for both absolute
magnitudes and distance measures. All distances are measured using
comoving coordinates.

\section{Data} \label{sec.data}

We use Sloan Digital Sky Survey (SDSS) data to determine the galactic
environment of quasars and galaxies. The SDSS
\citep{york_2000,stoughton_2002,pier_2003,Abazajian_2003,gunn_2005} is
conducting an imaging survey of $10^4$ square degrees of the sky in
five bands ($u, g, r, i, z$) \citep{fukugita_1996,gunn_1998}, followed
by a spectroscopic multi-fiber survey of the brightest $10^6$ galaxies
and $10^5$ quasars. The spectroscopic targets are selected from the
high quality imaging data using well-defined selection criteria
\citep{lupton_2001,hogg_2001,strauss_2002,richards_2002}. The
drift-scan imaging survey reaches a limiting magnitude of $r < 23$
\citep{fukugita_1996,gunn_1998,lupton_2001}. The main spectroscopic
survey targets galaxies to $r < 17.7$, with a median redshift of $z
\sim0.15$ and a tail reaching $z \sim0.4$ (Strauss et al 2002). The
spectroscopic survey of quasars , with $i < $19, reaches quasar
redshifts out to $z \sim5.4$. For more details on the SDSS see the
above references.

The high quality imaging and spectroscopic survey of quasars and
galaxies provides a unique data set for studying the environment of
quasars and comparing it directly with the environment of galaxies. To
do so, we use the third data release (DR3) of the SDSS spectroscopic
sample of quasars \citep{schneider_2005}, selecting all
spectroscopic quasars with redshift $z \le 0.4$ and $i$-band
Galactic extinction corrected and k-corrected magnitude $-24.2 \le M_i
\le -22.0$ \citep{richards_2002}. We set an upper limit of $-24.2$ on
the luminosity to avoid bright objects that may interfere with
counting nearby galaxies. After applying masks for missing fields (see
the end of this section), a sample of 2028 $z \le 0.4$ quasars is
used, covering an area of approximately 4000 deg$^2$. In addition to
the quasars, we use a sample of spectroscopic galaxies as targets in
our analysis so that we may compare the environment of quasars with
that of galaxies. The spectroscopic galaxy sample used for comparison
is the NYU-LSS sample 12 \citep{blanton_2003a,blanton_2003b}, which is
comprised of a complete spectroscopic sample of galaxies to
$i\,=\,18.5$ corrected for both Galactic extinction and k-correction,
with redshifts in the range $z~\sim0.001$ to $z~\sim0.4$. After
applying masks and limiting the galaxy redshift to $0.08 \le z \le
0.4$ (since there are no quasars with $z < 0.08$), a sample of
$\sim10^5$ spectroscopic galaxies is available over a $2230\,$ deg$^2$
area (mostly overlapping the quasar area). Our galaxy sample has a
median redshift of 0.13 and a median magnitude of $M_i = -21.3$, and
our quasar sample has a median redshift of 0.32 and a median magnitude
of $M_i = -22.5$.

The environment of the above targets - spectroscopic quasars and
spectroscopic galaxies - is then determined using the photometric
galaxies from DR3 of the SDSS imaging survey. We use a sample of over
10 million galaxies with magnitude in the range $14 \le i \le 21$. For
further comparison, we also repeat the environment analysis at
approximately $10^3$ random positions per target using the same method
and background sample of photometric galaxies.

All samples were corrected using the same masks, which remove missing
fields, missing stripes, and regions where the stripe boundaries
extend beyond the extent of the photometric galaxy survey. These masks
were created with SDSSpixel, a pixelization scheme routinely applied
to the SDSS data\footnote{See
http://lahmu.phyast.pitt.edu/\~{}scranton/SDSSPix/ for information on
SDSSpixel}. We also remove all targets (i.e., quasars, spectroscopic
galaxies, and random points) that are closer than $1\,$Mpc to the
boundary or to a mask in the photometric sample in order to ensure
that all targets have a complete field of photometric galaxies within
the scales of interest.
 
\section{Analysis} \label{sec.analysis}

We study the environment of the spectroscopic targets (quasars and
galaxies) by determining the number of photometric galaxies within
different projected radii from the quasars, from $25\,$kpc to $1\,$Mpc
($h=0.7$). In order to normalize the density of photometric galaxies,
we also estimate the density of photometric galaxies around a large
number of random positions in the survey area.  This method allows a
direct comparison of the observed density around quasars and around
galaxies with that found around random positions using the same
observed distribution of photometric galaxies. This latter comparison
yields normalized overdensities, i.e., observed density over random
density, for both the quasars and the spectroscopic galaxies. This
technique then allows a direct comparison of the density of galaxies
around spectroscopic targets.

Throughout the analysis, we treat all of these targets - quasars,
spectroscopic galaxies, and random points - in exactly the same manner
in order to provide a direct and straightforward comparison between
the environment of quasars, galaxies, and random positions, and help
minimize potential biases. For each of the targets, we determine the
number of photometric galaxies within projected comoving radius bins
from $25\,$kpc to $1\,$Mpc ($h=0.7$). The innermost $25\,$kpc
(15.3\arcsec to 3.3\arcsec over our redshift range) is removed in all
density estimations in order to avoid deblending issues at these small
separations. The number of photometric galaxies observed around
quasars and around spectroscopic galaxies, $\ngq$ and $\ngg$
respectively, is divided by the same number found around random
points, $\ngr$; these overdensities, $\ngq/\ngr$ and $\ngg/\ngr$, are
determined for each of the radii specified above. In order to account
for the different redshift distributions of the $z \le 0.4$ targets
(quasars and galaxies), the overdensities are determined for each
individual quasar or galaxy using the mean $\ngr$ appropriate for that
target's redshift. All overdensities are then averaged in the relevant
redshift bins, and are investigated as a function of radius,
luminosity, and redshift. Our environment estimator is less
sensitive to faint galaxies at high redshift, but it is still
informative as we are interested in an excess in quasar environment
density relative to that of galaxies. At low redshift, the average
number of photometric galaxies around quasars ranges from a few
galaxies within $100\,$kpc (typically twice the number found around
random points) to $\sim100\,$ galaxies within $1\,$Mpc.

In order to estimate the density errors and the correlations between
different scales, we have generated $10^5$ bootstrap samples of the
spectroscopic quasar, galaxy and random samples, and measured the
standard deviation among different realizations. As the counts of
photometric galaxies are dominated by projection effects, i.e. objects
uncorrelated to the spectroscopic targets, the error bars are close to
Poisson errors, especially on large-scales.

\section{Results} \label{sec.results}

\begin{figure}
\includegraphics[width=.49\textwidth]{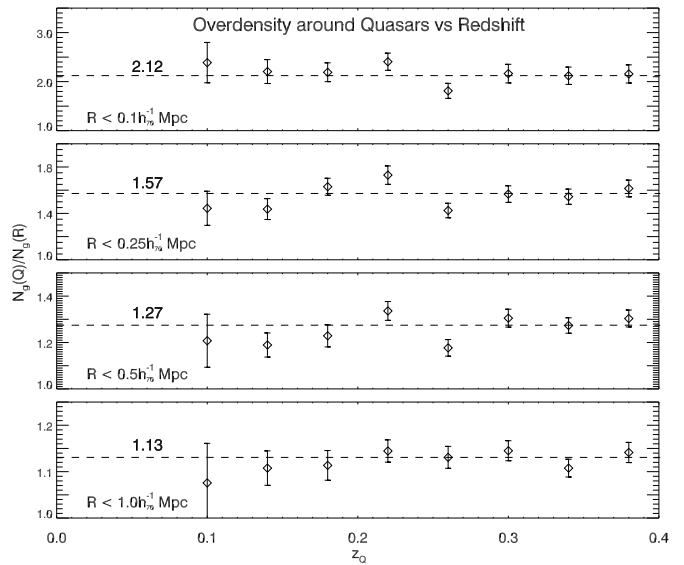}
\caption{Integrated overdensity (from $25\,$kpc to 0.1, 0.25, 0.5, and
$1.0\,$Mpc) around quasars ($-24.2 \le M_i \le -22$) as a function of
redshift. The dashed line indicates the average quasar
overdensity. All error bars are bootstrap errors.}
\label{f.z}
\end{figure}

Our analysis produces normalized overdensities around quasars and
around spectroscopic galaxies. We note that the normalized
densities we use are defined as ratios of the galaxy counts around
each target relative to that around random points (e.g. $\ngq/\ngr$);
a ratio of unity implies no overdensity, i.e. the same density around
quasars as around random positions. Another common definition of
overdensity relative to random can be calculated using
$\ngq/\ngr\,-\,1$. This can be directly obtained from the $\ngq/\ngr$
ratios provided below.

The quasar overdensity
is presented as a function of redshift in Figure~\ref{f.z} within each
of our four standard radii. The mean overdensity around quasars is
shown by the dashed lines. The overdensity is 2.12 within $0.1\,$Mpc
of the quasars; i.e., the density is 2.12 times larger than the
density around random points. The mean overdensity decreases
monotonically with radius; it is 1.57 within $0.25\,$Mpc, 1.27 within
$0.5\,$Mpc, and 1.13 within $1.0\,$Mpc of the quasars. This
overdensity refers to the mean of all quasars with $-24.2 \le M_i \le
-22$. The excess photometric galaxies refers to galaxies within the
magnitude range $14 \le i \le 21$ (Section~\ref{sec.data}). The mean
galaxy overdensity around quasars is independent of redshift for $z
\le 0.4$ (Figure~\ref{f.z}).

\begin{figure}
\includegraphics[width=.49\textwidth]{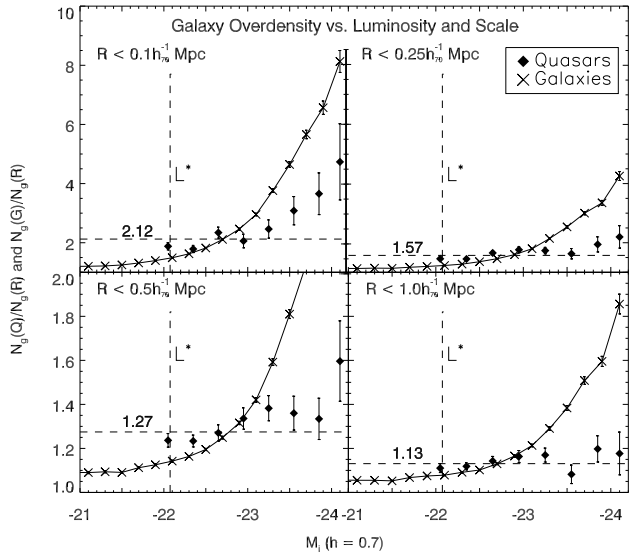}
\caption{Integrated overdensity as a function of quasar or galaxy
magnitude for four radii. The data are for the integrated redshift
range $0.08 \le z \le 0.4$. The horizontal dashed line indicates the
average quasar overdensity. The vertical dashed line indicates the
magnitude of $L^*$ galaxies, which we use as a standard reference for
comparison with the clustering around quasars. We see that quasars on
average inhabit higher local density regions than do $L^*$ galaxies;
their local environments are comparable to those of $\sim2L^*$
galaxies.}
\label{f.over}
\end{figure}

In Figure~\ref{f.over} we present the overdensity as a function of
luminosity for quasars and for spectroscopic galaxies within radii of
0.1, 0.25, 0.5, and $1\,$Mpc ($h=0.7$). We find that, at all radii,
the overdensity around galaxies (solid line) increases with galaxy
luminosity.  This is expected as brighter galaxies are located, on
average, in higher density regions \citep[e.g.,][]{davis_1988,
hamilton_1988, white_1988, zehavi_2004, eisenstein_2005}.  The galaxy
overdensity is greatest on small scales ($0.1\,$Mpc and closer), and
decreases on larger scales, as expected. The luminosity-overdensity
trend is steeper on small scales than on large scales. The overdensity
around $L^*$ galaxies is $1.51\pm0.01$ times larger than random within
a radius of $0.1\,$Mpc; the overdensity is nearly doubled, to
$2.95\pm0.05$, for galaxies that are brighter by 1 magnitude. 

The quasar overdensity in Figure~\ref{f.over} shows an increase with
quasar luminosity on the smallest scales, but only for the most
luminous quasars. The trend is considerably weaker than for galaxies
and becomes negligible, with nearly no dependence on quasar luminosity
by $\sim0.5-1\,$Mpc scales as well as for quasars with lower
luminosities (Figure~\ref{f.over}). This indicates that there is
little or no correlation between quasar activity and environment on
these larger scales, and that quasar activity is typically triggered
by an interaction with a neighbor present within $\sim100\,$kpc from
the quasar host galaxy. The mean overdensity around quasars,
$<\ngq/\ngr>$, for all $z \le 0.4$ quasars brighter than $M_i \le -22$
is indicated by the horizontal dashed line in Figure~\ref{f.over}. The
mean overdensity decreases with radius. The results are summarized in
Table~1.

\setcounter{table}{0}
\begin{table}
\caption{Mean Galaxy Overdensity around Quasars ($-24.2 \le M_i \le
  -22.0$, $z \le 0.4$) and around $L^*$ Galaxies.}
\begin{tabular}{cccc}
\hline \hline $R_{max}$ (Mpc)&
$\frac{\ngq}{\ngr}$&$\frac{\ngl}{\ngr}$&$\frac{\ngq/\ngr}{\ngl/\ngr}$\\
\hline 0.10 & $2.12\pm0.08$ & $1.507\pm0.010$ & $1.41\pm0.06$\\ 0.25 &
$1.57\pm0.03$ & $1.235\pm0.004$ & $1.27\pm0.03$\\ 0.50 & $1.27\pm0.02$
& $1.144\pm0.003$ & $1.11\pm0.02$\\ 1.00 & $1.13\pm0.01$ &
$1.082\pm0.002$ & $1.05\pm0.01$\\ \hline
\end{tabular}
\end{table}

Figure~\ref{f.over} allows us to compare the environment of quasars
with the environment of galaxies of different magnitudes. We use the
average overdensity around $L^*$ galaxies as a standard by which to
measure the quasar environment. Since quasar luminosity is clearly
physically unrelated to the galactic luminosity, it should be noted
that $L^*$ galaxies are used only as an frame of reference for
comparing the relative environments. We find the mean overdensity
around quasars to be larger than around $L^*$ galaxies (shown by the
vertical dashed line in Figure~\ref{f.over}) on all scales
$<1\,$Mpc. The mean quasar overdensity is larger than the overdensity
around $L^*$ galaxies by factors that range from $1.41\pm0.06$ within
$0.1\,$Mpc, to $1.27\pm0.03$ within $0.25\,$Mpc, $1.11\pm0.02$ within
$0.5\,$Mpc, and $1.05\pm0.01$ within $1\,$Mpc (Table~1). (Using the
alternate overdensity definition,
$[\ngq/\ngr\,-\,1]\,/\,[\ngl/\ngr\,-\,1]$, we find an excess ratio of
$2.2\pm0.16$ within $0.1\,$Mpc decreasing to $1.6\pm0.13$ within
$1\,$Mpc. The interpretation is, of course, the same.) The local
density enhancement around quasars is similar to the local enhancement
around $\sim2L^*$ galaxies. The overdensity around quasars relative
to $L^*$ increases to a factor of $2.8\pm0.56$ closest to the quasars,
at $\sim40\,$kpc (see below). On scales larger than $1\,$Mpc, the
quasar overdensity becomes similar to the environment of $L^*$
galaxies. These results indicate that quasars are located in higher
density regions than are $L^*$ galaxies, but that the overdensity
exists mostly in regions very close to the quasars
($\lesssim0.1\,$Mpc), and is thus a very local excess. This local
excess of galaxies near quasars likely plays an important role in
triggering the quasar activity through mergers and other interactions.

\begin{figure}
\includegraphics[width=.49\textwidth]{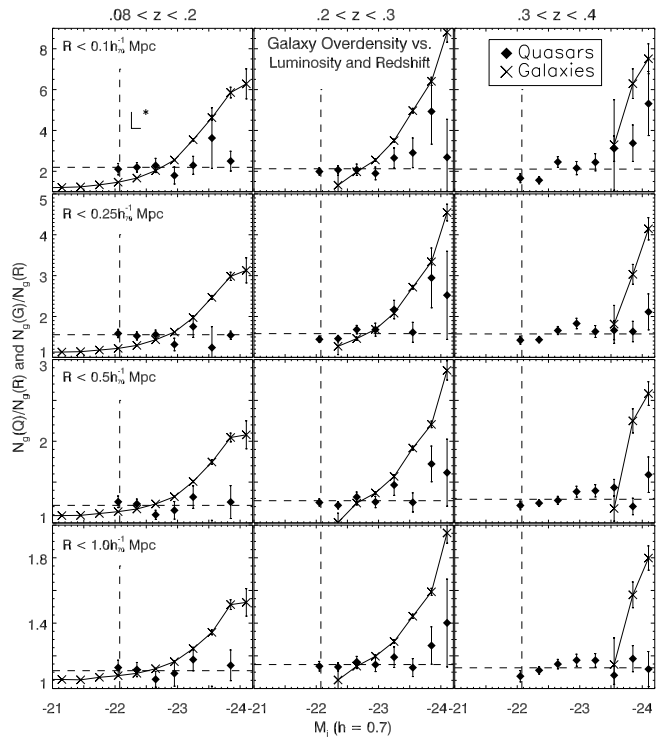}
\caption{Integrated galaxy overdensity around quasars and around
galaxies for our four radii (rows) as a function of seed magnitude for
three redshift bins (columns).}\label{f.overz}
\end{figure}

\begin{figure}
\includegraphics[width=.49\textwidth]{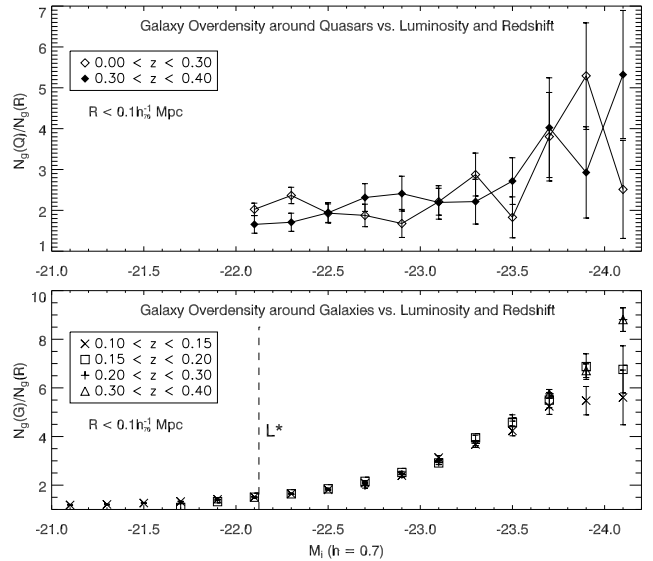}
\caption{Integrated overdensity (to $0.1\,$Mpc) as a function of seed
magnitude. Both quasars (top) and galaxies (bottom) have been
separated into redshift bins. Note that there are many overlying
points in the galaxy overdensity plot.}
\label{f.overz2}
\end{figure}

These results are found to be only weakly dependent on redshift in the
$z < 0.4$ range studied here. This is shown in Figure~\ref{f.overz},
where the results, similar to Figure~\ref{f.over}, are presented for
different redshift ranges.  As can be seen, the trend of overdensity
with quasar and galaxy luminosity remains consistent within the
statistical uncertainties and no significant trend can be detected as
a function of redshift. The redshift independence is further
illustrated in Figure~\ref{f.overz2}, where the overdensity around
quasars and the overdensity around galaxies are plotted as a function
of luminosity for several redshift bins. The results show a redshift
independent overdensity signal for both quasars and for galaxies. (The
highest luminosity galaxies, which do show some evolution, do not
affect our conclusions as we compare our quasars only to $L^*$ galaxies.)
Our results for the overdensities and the comparison with the
environment of $L^*$ galaxies are unaffected by redshift.

\begin{figure}
\includegraphics[width=.49\textwidth]{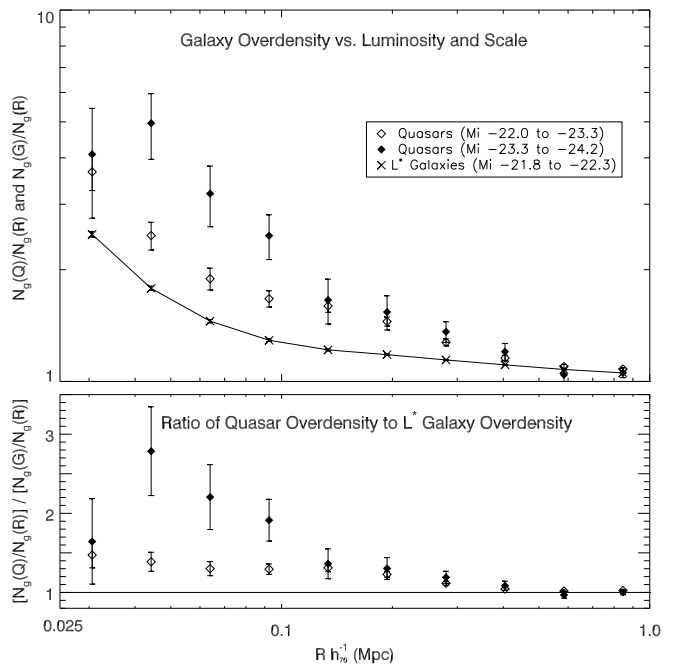}
\caption{Top: Overdensity as a function of comoving radius for bright
quasars, dim quasars, and $L^*$ galaxies. $L^*$ galaxies were defined
to be within a quarter of a magnitude of the canonical value of $M_i =
-22.125$. Bottom: Ratio of quasar overdensity to $L^*$ galaxy
overdensity for both bright and dim quasars. The anomalously low point
for the brightest quasars at $30\,$kpc is the result of the brightest
quasars extending beyond our $25\,$kpc central cut.}
\label{f.corr}
\end{figure}

The scale dependence of the galaxy overdensity around spectroscopic
quasars and galaxies is presented as a function of comoving distance
from the target object in Figure~\ref{f.corr}. The quasar
overdensities are shown for both the brightest quasars ($M_i = -23.3$
to $-24.2$) and for fainter quasars ($M_i = -22$ to $-23.3$). We find
that the quasar overdensities are larger than those of $L^*$ galaxies
at all radii less than $\sim0.5\,$Mpc, but the overdensity increases
substantially on smaller scales. The overdensities around the most
luminous quasars are larger than around fainter quasars, mostly on
small scales ($\le\,0.1\,$Mpc). The lower panel of Figure~\ref{f.corr}
presents the ratio of quasar overdensity to that of $L^*$ galaxy
overdensity, as a function of scale. These data illustrate the
transition between the large scales, where quasars and $L^*$ galaxies
inhabit similar environments, and the small scales, where quasars are
located in higher overdensity regions than are $L^*$ galaxies. On
these small scales, the quasar overdensity is comparable to that
around $\sim2L^*$ galaxies (or brighter, for the most luminous
quasars). Our results on scales greater than $0.5\,$Mpc are
consistent with the recent SDSS and 2dF research on the large scale
quasar-galaxy clustering \citep[e.g.,][]{smith_1995, croom_1999,
croom_2003, wake_2004}, while our results on smaller scales are
consistent with the early work discussed in Section~\ref{sec.intro}
\citep[e.g.,][]{shaver_1988, shanks_1988, chu_1988, chu_1989,
crampton_1989} as well as with recent quasar-quasar pair studies
\citep{hennawi_2005}. This difference in the relative quasar
environment between small ($\le\,0.5\,$Mpc) and large ($\ge\,1\,$Mpc)
scales explains some of the previously contradictory results discussed
in Section~\ref{sec.intro}. These results suggest a picture in which
quasars reside in, on average, galaxies with a luminosity comparable
to $\sim L^*$, but with a local excess of neighbors within
$\sim0.1\,-\,0.5\,$Mpc. This local excess is likely associated with
triggering quasar activity.

The different galaxy densities observed for the bright and faint
quasars cannot be due to gravitational lensing. The dark matter
distribution in and around galaxies induce gravitational lensing
effects which locally enlarge the sky solid angle and magnify the
images of background objects. This effect, the magnification bias, can
increase or decrease the density of distant sources around foreground
galaxies depending on the variation of the number of sources as a
function of magnitude \citep{narayan_1989}. It can thus create
correlations between source luminosity and foreground galaxy
density. However, the amplitude of this effect is expected to be only
at the percent level \citep{menard_2002, jain_2003} and recent
observations with the SDSS have confirmed these predictions
\citep{scranton_2005}. This suggests that gravitational lensing does
not significantly affect our density estimations.

\section{Conclusions} \label{sec.conclusions}

We use the SDSS data to investigate the local environment of quasars
and compare it with the environment around galaxies and around random
positions. This study provides a direct comparison between the
environment of quasars and galaxies.

We use bright quasars with $-24.2 \le M_i \le -22$ ($h=0.7$) and
redshift $z \le 0.4$ (2028 quasars over $4000\,$deg$^2$) to study the
density of photometric galaxies with $i = 14$ to 21 (sample of
$\sim10^7$ galaxies) located within $25\,$kpc to $1\,$Mpc ($h=0.7$) of
the parent quasars. We compare the results with the same analysis
carried out at random positions in the SDSS survey, $\ngr$; this
yields the galaxy overdensity, over random, around quasars, i.e.,
$\ngq/\ngr$. We investigate this overdensity as a function of quasar
redshift and luminosity. The same analysis is then repeated for
determining the overdensity around $10^5$ spectroscopic galaxies in
the SDSS data ($i \le 18.5$) in the same redshift range ($z = 0.08$ to
$0.4$). This allows a direct comparison of the quasar overdensity to
the overdensity around galaxies. The overdensities are studied as a
function of scale, luminosity, and redshift. This comparison of quasar
environment with that of galaxies provides a self-consistent
comparison of the host environments.

Our results are summarized below.

1. At all radii, from $25\,$kpc to $\sim1\,$Mpc, quasars are located
   in higher density environments than are $L^*$ galaxies. The
   overdensity around quasars relative to that of $L^*$ galaxies
   increases with decreasing scale: the overdensity is greatest
   closest to the quasars. At a distance of $40\,$kpc of the quasars,
   the mean overdensity for the brightest quasars ($-24.2 \le Mi \le
   -23.3$, $z \le 0.4$) is nearly a factor of three times larger than
   the overdensity around $L^*$ galaxies. The mean overdensity around
   the brightest quasars relative to that of $L^*$ galaxies decreases
   with increasing scale to a value of 2.2 within $0.1\,$Mpc, 1.2 at
   $0.3\,$Mpc, and approaches unity at $\sim0.5 - 1\,$Mpc. On these
   larger scales, quasars reside in environments similar to those of
   $L^*$ galaxies.

2. The brightest quasars are found to be located in higher overdensity
   regions than are fainter quasars, especially at small separations
   ($< 0.1\,$Mpc). On larger scales, bright and faint quasars live in
   similarly dense environments.

3. The mean overdensity around quasars 0
($<~\ngq/\ngr~> = 2.12\pm0.08$
   within $0.1\,$Mpc, decreasing to $1.13\pm0.01$ within $1\,$Mpc) is
   independent of redshift for $z \le 0.4$. The mean overdensity is
   also independent of luminosity except for the brightest quasars,
   which are located in higher density environments. This dependence
   of quasar environment on luminosity, showing enhancement only for
   the most luminous quasars, is consistent with recent galaxy merging
   models of quasars \citep{hopkins_2005}.

4. The enhanced mean overdensity around quasars is observed to be a
   local phenomenon, affecting mostly the $\sim$ $0.1\,$Mpc region
   closest to the quasars. On these scales, very close to the quasars,
   the high overdensity of galaxies likely affects the formation and
   triggering of the quasar activity through mergers and other
   interactions. On scales of $\sim 1\,$Mpc, the quasars inhabit
   similar environments to those of normal $L^*$ galaxies.

\section{Acknowledgements} \label{sec.acknowledgements}

GTR acknowledges support from a Gordon and Betty Moore Fellowship in
data intensive sciences.

Funding for the creation and distribution of the SDSS Archive has been
provided by the Alfred P. Sloan Foundation, the Participating
Institutions, the National Aeronautics and Space Administration, the
National Science Foundation, the U.S. Department of Energy, the
Japanese Monbukagakusho, and the Max Planck Society. The SDSS Web site
is http://www.sdss.org/.

The SDSS is managed by the Astrophysical Research Consortium (ARC) for
the Participating Institutions. The Participating Institutions are The
University of Chicago, Fermilab, the Institute for Advanced Study, the
Japan Participation Group, The Johns Hopkins University, the Korean
Scientist Group, Los Alamos National Laboratory, the
Max-Planck-Institute for Astronomy (MPIA), the Max-Planck-Institute
for Astrophysics (MPA), New Mexico State University, University of
Pittsburgh, University of Portsmouth, Princeton University, the United
States Naval Observatory, and the University of Washington.


\end{document}